\documentclass[prl,twocolumn,preprintnumbers,amsmath,amssymb,epsfig,superscriptaddress]{revtex4}

\usepackage{epsfig}
\usepackage{mathrsfs}
\usepackage{amssymb}

\usepackage{graphicx}
\usepackage{dcolumn}
\usepackage{bm}

\newcommand{\set}[1] {\{ #1 \} }

\begin{document}

\title{Reconstructing nonlinearities with intermodulation spectroscopy}

\address{}

\author{Carsten Hutter}
\affiliation{Department of Physics, Stockholm University}
\affiliation{Nanostructure Physics, Royal Institute of Technology (KTH)\\
AlbaNova, SE--106 91 Stockholm, Sweden}

\author{Daniel Platz}
\affiliation{Nanostructure Physics, Royal Institute of Technology (KTH)\\
AlbaNova, SE--106 91 Stockholm, Sweden}

\author{E. A. Thol{\'e}n}
\affiliation{Nanostructure Physics, Royal Institute of Technology (KTH)\\
AlbaNova, SE--106 91 Stockholm, Sweden}

\author{T. H. Hansson}
\affiliation{Department of Physics, Stockholm University}

\author{D. B. Haviland}
\affiliation{Nanostructure Physics, Royal Institute of Technology (KTH)\\
AlbaNova, SE--106 91 Stockholm, Sweden}

\date{\today}

\begin{abstract}
We describe a method of analysis which allows for reconstructing the nonlinear disturbance of a high $Q$ harmonic oscillator.  When the oscillator is driven with two or more frequencies, the nonlinearity causes intermodulation of the drives, resulting in a complicated spectral response.  Analysis of this spectrum allows one to approximate the nonlinearity.  The method, which is generally applicable to measurements based on resonant detection, increases the information content of the measurement without requiring large detection bandwidth, and optimally uses the enhanced sensitivity near resonance to extract information and minimize error due to detector noise.
\end{abstract}

                             
\maketitle

A harmonic oscillator with high quality factor $Q$, i.e., large stored energy in relation to the energy lost in each cycle, is a very useful tool for precision measurement.  When the oscillator interacts with an object of interest, the first order correction to the linear response is a shift of the resonant frequency $\omega_0$, which results in a large change of the amplitude and phase of the response near resonance. Resonant detection is employed in a wide variety of measurements, and particularly in nanoscience, where examples include the radio frequency single electron transistor \cite{science1998schoelkopf}, dispersive readout of superconducting qubits 
\cite{prl2005wallraff}, measuring the deflection of a nanomechanical oscillator \cite{naturePhysics2008regal}, and nanoscale imaging of surfaces with dynamic atomic force microscopy (AFM) \cite{ssr2002garcia}.   
In this letter we describe a method of resonant detection which exploits the nonlinear phenomenon of intermodulation, to generate a spectral response near resonance that enhances the information content of the measurement.   We provide a general theoretical framework for analyzing the intermodulation spectrum and for reconstructing the nonlinearity. 

The sensitivity of resonant detection can be understood from the large transfer gain, $G(\omega)$, of a high-$Q$ oscillator.  In the narrow frequency band  $B=\omega_0/Q$ centered at resonance, the oscillator gives large response to a small stimulus.  When the linear oscillator interacts with the object of interest, the equation of motion becomes nonlinear and higher harmonics of the drive frequency will appear in the response.  One could in principle analyze the amplitude and phase of these higher harmonics to extract the nonlinearity.  The problem with this approach is: (i) it requires a very large detection bandwidth, $B_{\rm D}=n \omega_0$ to get $n$ harmonics and (ii) the transfer gain for harmonics is far less than one, falling off as $1/\omega^2$ above resonance, so the response at higher harmonics is often buried in the detector noise.  In the field of AFM, the latter problem was addressed by coupling more oscillators, utilizing additional torsional \cite{natureNanotech2007sahin} and flexural \cite{sensors2004sahin,  prl2009xu} eigenmodes of the cantilever.  In this context, different types of multi-tone excitation schemes have been put forward as a means of extracting qualitative information about the nonlinearity \cite{apl2004rodriguez,nanotechnology2007jesse,prl2008lozano}.  

\begin{figure}
\includegraphics[width=7cm]{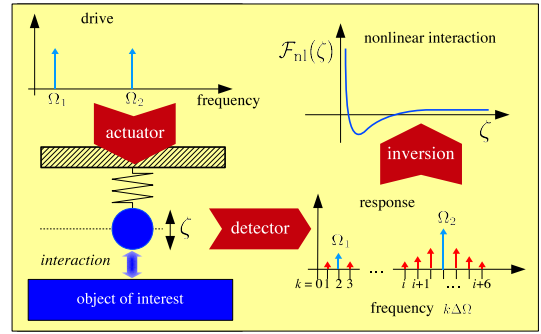} 
\caption{(color online). Intermodulation spectroscopy: two pure tones drive the resonator which interacts with an object of interest.  The detected response spectrum consisting of many intermodulation products can be inverted to reconstruct the nonlinear interaction.}
\label{fig:schematic}
\end{figure}

An alternative approach, sketched in Fig.~\ref{fig:schematic}, is to drive the oscillator with two pure tones $\omega_1$ and $\omega_2$, so as to produce intermodulation products (IMPs) or ``mixing products'' which occur at the frequencies $n\omega_1 + m \omega_2$ ($n$ and $m$ integers), where the order of the IMP is defined by $\vert n \vert + \vert m \vert$.  The drive frequencies can be chosen in a variety of ways, with the objective of creating many IMPs near resonance where the sensitivity is enhanced. These IMPs measured near resonance represent a partial spectral response of the system from which one would like to reconstruct the nonlinearity.

\begin{figure}
\includegraphics[width=8cm]{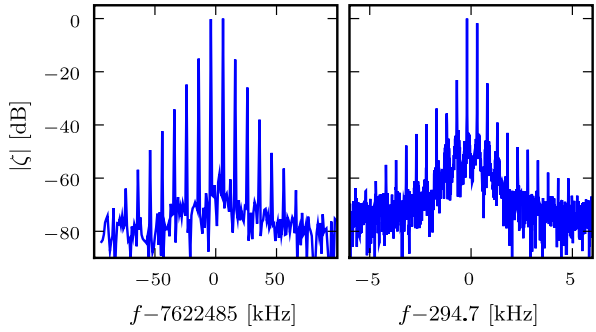} 
\caption{(color online). Measured intermodulation spectra taken from a nonlinear superconducting resonator \cite{physicaScripta2009tholen} (left panel) and an AFM cantilever interacting with a surface \cite{proceedings2009platz}
(right panel).  In these experiments the two central peaks with largest response are the drive tones.  The amplitude in dB is normalized to the maximum response. }
\label{fig:spectra}
\end{figure}

Our work is motivated by the desire to understand the information content of intermodulation spectra, as measured in superconducting microresonators at GHz frequencies \cite{apl2007tholen,physicaScripta2009tholen}, 
and in AFM cantilevers oscillating at several hundred kHz while interacting with a surface \cite{apl2008platz, proceedings2009platz} (see Fig.~\ref{fig:spectra}).  However, we wish to stress that the reconstruction algorithm described here can be applied to a wide variety of experiments which exploit resonant detection.  To test the reconstruction algorithm, we use simulated data obtained by numerical integration of the equation of motion of the nonlinear oscillator, where, in contrast with experiment, we can work backward from the simulated intermodulation spectrum to the \emph{known} expansion coefficients of the nonlinearity, thus allowing a more
systematic study of errors.  

The equation of motion of a single-mode harmonic oscillator, which is driven by two pure tones, subject to viscous damping, and perturbed by a nonlinearity, reads:  
\begin{equation}
\label{eq:equationOfMotion}
\ddot{\zeta}+\frac{\dot{\zeta}}{Q}+\zeta =\mathcal{F}_{\rm
nl}(\zeta)+\mathcal{F}_1\cos(\Omega_1 \tau)+\mathcal{F}_2\cos(\Omega_2 \tau) \, .
\end{equation}
Here, $\Omega_{1,2}=\omega_{1,2}/\omega_0$, dots denote derivatives with respect to the dimensionless time $\tau=\omega_0 t$,  and the dimensionless coordinate $\zeta$ is the deviation of the oscillator from the equilibrium position, suitably normalized.  The dimensionless nonlinear ``force'' ${\mathcal F}_{\rm nl}(\zeta)$ is assumed conservative, with no explicit time-dependence.

While the motion of a damped, driven nonlinear system of the type (\ref{eq:equationOfMotion}) can in general be either regular or chaotic, 
we restrict ourselves here to nonlinearities and drive strengths which are weak enough, such that for commensurate drive frequencies with greatest common divisor $\Delta\Omega\equiv\mathrm{gcd}(\Omega_1,\Omega_2)$  the motion in the steady state is periodic in $2 \pi / \Delta\Omega$.
We may therefore expand the response in a discrete Fourier series, $\zeta(\tau)=\sum_k \zeta_k e^{ik\ \Delta\Omega\ \tau}$, where the expansion coefficients $\zeta_k$ are the complex numbers (amplitude and phase) which constitute the intermodulation spectrum.  With Fourier expansions of the nonlinear force $\mathcal{F}_{\rm nl}(\zeta(\tau))=\sum_k \mathcal{F}_{{\rm nl},k} e^{ik\ \Delta\Omega\ \tau}$, and the drive force,  $\mathcal{F}_{\rm d}(\tau)=\sum_k \mathcal{F}_{{\rm d},k} e^{ik\ \Delta\Omega\ \tau}$, the equation of motion (\ref{eq:equationOfMotion}) becomes
\begin{equation}
\label{eq:fourierEquation}
\zeta_k=   G_k ( \mathcal{F}_{{\rm d},k} +  \mathcal{F}_{{\rm nl},k} ) \equiv  \zeta_k^{(0)} +G_k \mathcal{F}_{{\rm nl},k}\ ,
\end{equation}
where $G_k=1/(1-k^2\Delta\Omega^2+ik\Delta\Omega/Q)$. 
Note that while (\ref{eq:fourierEquation}) is an exact equation, 
it is  not a solution of (\ref{eq:equationOfMotion}), since $\mathcal{F}_{{\rm nl},k}$ depends on all the $\zeta_k$ in a highly nonlinear way.  

To proceed, we represent the nonlinear force as a polynomial in $\zeta(\tau)$, 
 $\mathcal{F}_{\rm nl}(\zeta(\tau))=\sum_{j=1}^{\infty} g_{j} \zeta^j(\tau)$, which allows us to express the Fourier components of the {\it nonlinear} force $\mathcal{F}_{{\rm nl},k}$ as a {\it linear} combination of the polynomial coefficients, $g_j$:
\begin{equation}
\label{eq:Ftsk}
\mathcal{F}_{{\rm nl},k}=\sum_{j=1}^\infty H_{kj} g_j,
\end{equation}
where the matrix element $H_{kj}$ is the $k$th spectral component of the Fourier transform of $\zeta^j(\tau)$, alternatively written as $H_{k1}=\zeta_k$ and 
$H_{k,j+1}=\sum_{k'} \zeta_{k'} H_{k-k',j}$.

For a known nonlinear force, we immediately find a first approximation $\zeta_k^{(1)}$ of the spectrum, by evaluating $\mathcal{F}_{{\rm nl},k}$ in (\ref{eq:fourierEquation}) at the free solution 
$\zeta_k^{(0)}=G_k \mathcal{F}_{{\rm d},k}$.
In the next step we use $\zeta_k^{(1)}$ in the same way to obtain $\zeta_k^{(2)}$, etc. Each iteration generates new spectral response, but only at the intermodulation frequencies. If this procedure converges
to a solution $\zeta_k$, then our assumption of periodic orbits is certainly justified.  However, here we are interested in the inverse problem of finding the nonlinear force, starting from 
$\zeta_k^{\rm{exp}}$, which is measured with limited bandwidth
in the presence of noise.  Inspection of Eqs. (\ref{eq:Ftsk}) and (\ref{eq:fourierEquation}) reveals that one must invert the matrix $\mathbf{H}$ to find the expansion coefficients $g_j$ of the nonlinearity:  
\begin{equation}
\label{eq:inversion}
g_j = \sum_k ({\mathbf H}^{-1})_{jk}\frac{\zeta_k-\zeta^{(0)}_k }{G_k} \   \mbox{ for }\   j\in \set{1,...,j_{\rm max}  } \ .
\end{equation}

The inverse problem is generally complicated by the fact that the measured spectrum $\zeta_k^{\rm{exp}}$ has only a finite number $N_p$ of peaks that can be observed above the detector noise level, notably those close to the resonance frequency.  Nevertheless, with a proper intermodulation drive scheme one can achieve enough intermodulation peaks with good signal-to-noise ratio to construct a matrix $\mathbf{H}$, and calculate the coefficients $g_j$.  We use the following algorithm:

1. Choose the drive frequencies so that the frequency spacing of IMPs, $\Delta \Omega$, gives $N_p$ peaks in the measured intermodulation spectrum $\zeta_k^{\rm exp}$  at frequencies $k\Delta \Omega$ for $k\in \set{k_1, \dots, k_{N_p}}$, where $N_p$ is at least twice the number of expansion coefficients $j_{\rm max}$ desired in approximating the nonlinearity.\\
2. Define the intermodulation response $ {\zeta_k}$ to be  $\zeta_k=\zeta_k^{\rm exp}$ for $ k\in \set{k_1, \dots, k_{N_p}}$ and  $\zeta_k=0$ otherwise.\\
3. With the intermodulation response $ {\zeta_k}$, calculate the matrix $\mathbf{H}$, restricted to $j\le j_{\rm max}$ and $k\le k_{\rm max}\cdot j_{\rm max}$, 
where $k_{\rm max}={\rm max}\set{k_1, \dots, k_{N_p}}$.\\
4. Calculate the $j_{\rm max}$ expansion coefficients $g_j$  (restricted to be real) using Eq.~(\ref{eq:inversion}), with the inverse replaced by a pseudo inverse, the sum restricted to $k \in \set{k_1, \dots, k_{N_p}}$, and
$\mathbf{H}$ restricted to $ k\in \set{k_1, \dots, k_{N_p}}$ and to $j\le j_{\rm max}$.  
The pseudo inverse gives a least square fit of the $j_{\rm max}$ coefficients $g_j$ to the $N_p$ equations $\zeta_k= \zeta_k^{(0)} +G_k \sum_{j=1}^{j_{\rm max}}  H_{kj} g_j$, thus
reducing sensitivity to both measurement errors and systematic errors in the calculation of  $\mathbf{H}$.

We have tested this reconstruction algorithm using realistic parameters for two different experimental realizations. In each case, the response of the nonlinear oscillator was simulated by numerical integration of Eq.(\ref{eq:equationOfMotion}) using the Fortran solver DDASKR \cite{ddaskr}.  The output of this integrator was sampled appropriately and fast Fourier transformed to generate an intermodulation spectrum.  Tests were made to ensure that there was no Fourier leakage of the spectral peaks, so that the background level in the spectrum corresponded to the error tolerance set in the integrator.  
Detector noise was simulated by adding random gaussian noise $\Delta \zeta(\tau)$ to each time sample, before the Fourier transform.

As a first example we consider a superconducting coplanar wave-guide resonator with a Josephson tunnel junction or weak link, currently studied for low-noise amplification and the readout of quantum bits \cite{apl2007tholen,naturePhysics2008castellanos-beltran,epl2009abdo,arxiv2007boaknin,bertet}.  The classical dynamics of the resonator can be mapped to our model equation by using $\zeta=I/I_0$, where $I$ is the current in the center strip and $I_0$ the critical current.  The quality factor $Q$ can be engineered in a wide range, up to $10^6$. Typically a nonlinear kinetic inductance provides a nonlinear ``force'' which is constrained by $g_j=0$ for even $j$, because the inductance $d\mathcal{F}_\mathrm{nl}/d\zeta$ should not depend on the sign of the current $\zeta$. We can thus employ a drive scheme which is only sensitive to IMPs of odd order, with drive frequencies closely spaced near resonance, $\Delta \Omega\equiv\Omega_2-\Omega_1\ll1$. This drive scheme was used for the experimental spectra shown in Fig. 2. 
For weak drive, spectra are well described by the lowest nonlinearity of Kerr-type, $g_3$, but for strong drive, higher nonlinear coefficients ($g_5$,$g_7$, ...) play a role.

In Tab.~\ref{tab:noise} we give the percent error in the coefficients $g_j^{\rm{recon}}$ which were reconstructed from data that was simulated using three coefficients ${g_3,g_5,g_7}$.  Results are shown for two different simulated noise levels (average of 1000 simulations) and two different quality factors.  In the absence of noise, $\Delta \zeta =0$, the small errors in the reconstruction are explained by the fact that we only used $20$ closely spaced spectral peaks around the resonance, leading to a slight miscalculation of the matrix $\mathbf H$.  These systematic errors, which are typically small,  can be reduced by using an improved reconstruction algorithm which estimates the unmeasured spectral peaks to achieve a self-consistent solution of Eq.(\ref{eq:fourierEquation}).  In the presence of noise, reconstruction from the intermodulation spectrum gives excellent results for all coefficients.  For comparison, we applied the reconstruction algorithm to a simulated spectrum of the same number of harmonics, generated with a single drive having strength $\mathcal{F}_{\rm d}=\mathcal{F}_{\rm d1}+\mathcal{F}_{\rm d2}$, for the same parameters given in Tab.~\ref{tab:noise}.  In the presence of noise, the spectrum of harmonics was unable to reconstruct the coefficients $g_5$ and $g_7$.
This nicely illustrates the advantage of resonant detection in an intermodulation scheme.

\begin{table}
\begin{centering}
\begin{tabular}{|l|r|r|r|r|r|r|}
\hline
&\multicolumn{3}{|c|}{$Q=50$}&\multicolumn{3}{|c|}{$Q=500$}\\ \cline{2-7}
&$\Delta g_3$&$\Delta g_5$&$\Delta g_7$&$\Delta g_3$&$\Delta g_5$&$\Delta g_7$\\ \hline
$\Delta\zeta=0$&0.04&0.16&0.20&0.00&0.02&0.00\\ \hline
$\Delta\zeta=10^{-3}$&0.10&0.28&0.26&0.03&0.19&0.33\\ \hline
$\Delta\zeta=10^{-2}$&0.87&2.43&1.90&0.32&1.96&3.35\\ \hline
\end{tabular}
\par\end{centering}
\vspace{-0.2cm}
\caption{ 
\label{tab:noise}
Relative errors $\Delta g_j\equiv |(g_j-g_j^{\rm{recon}})/g_j|$ in \%, for different strengths of random Gaussian noise in time, $\Delta \zeta(\tau)$, relative to the maximum response amplitude. Entries 0.00 mark relative errors smaller than $5\cdot 10^{-5}$. To simulate data, we used $g_1=0, g_3=10^{-3}$, $g_5=-10^{-4}$, $g_7=10^{-5}$.  These four coefficients were free parameters for the reconstruction. The two drives were centered on resonance.  For the case $Q=50$, $\Delta \Omega=1/499.5$ and $\mathcal{F}_{\rm d1}=\mathcal{F}_{\rm d2}=0.03$. For the case $Q=500$,  $\Delta\Omega=1/4999.5$ and $\mathcal{F}_{\rm d1}=\mathcal{F}_{\rm d2}=0.003$.
In both cases, $2\cdot10^5$ time samples were used in the Fourier transform.
\vspace{-0.5cm}
} 
\end{table}

As a second example, we consider an application which contains coefficients $g_j$ of both odd and even order.  In dynamic atomic force microscopy (AFM), the resonator is a cantilever oscillating about its equilibrium position with dimensionless amplitude $\zeta = z/z_s$, effective mass $m$, and stiffness $k=m\,\omega_0^2$.  The nonlinearity is provided by the tip-surface force, $F_{\rm nl}=F_{\rm ts}(\zeta)$, which we wish to determine.  Intermodulation spectroscopy can be performed while scanning over a surface, measuring the dominant IMPs at each image point \cite{apl2008platz, proceedings2009platz}. The inversion algorithm described here can be used to rapidly determine the force-distance curve at each image pixel, while scanning at normal speeds for dynamic AFM, one of the major objectives of AFM development \cite{pnas2002stark,natureNanotech2007sahin}.

The nonlinear force typically encountered in AFM experiments has an attractive region close to a repulsive surface.  We restrict our study to conservative forces, derivable from a potential function, $F_{\rm ts}=-dU/dz$, and desire to reconstruct a polynomial approximation of the dimensionless force $\mathcal{F}_{\rm ts}(\zeta)\equiv F_{\rm ts}(z+z_0)/(z_s k)$, where $z_0$ is the probe height above the surface in the absence of drive signals, an experimentally tunable parameter.  
For a Morse potential, this force has the functional form $\mathcal{F}_{\rm ts}=2 \alpha (e^{-2 (\zeta-\sigma) /\lambda}-e^{-(\zeta-\sigma) /\lambda})$, which is described by 
the location of the potential minimum, $\sigma$, its width, $\lambda$, and an overall scale factor, $\alpha$ \cite{rmp2003giessibl}.  We used this nonlinear force to simulate data with a quality factor $Q=50$ and two drives centered around resonance.  Because this drive scheme generates near resonance only IMPs of odd order, intermodulation peaks around twice the resonance frequency were needed to gain information about the coefficients $g_j$ with even $j$.  
20 peaks around resonance and 10 peaks around $2\omega_0$ were taken for the reconstruction.  All other spectral information was set to zero.  
\begin{figure}
\includegraphics[width=8cm]{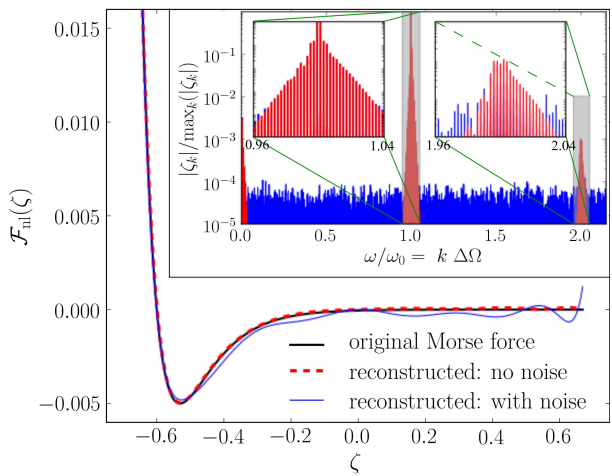} 
\vspace{-0.3cm}
\caption{(color online). The original force $F_{\rm ts}(\zeta)$ used in the simulation, a plot of the polynomial described by the reconstructed coefficients $g_j$, and the latter in the presence of noise.   Inset: The spectrum with noise (blue) and overlaid without noise (red). Parameters: $\alpha=0.01$, $\sigma=-0.6$, $\lambda=0.1$, $\Delta \Omega=1/499.5$, and $\mathcal{F}_{\rm d1}=\mathcal{F}_{\rm d2}=0.007$.}
\label{fig:afm_recon}
\vspace{-0.3cm}
\end{figure}

The reconstruction algorithm was run using the amplitude and phase of these $N_p=30$ peaks, and $j_{\rm{max}}=15$ coefficients $g_j$ were taken for the polynomial representation of the nonlinear force.  The result of the reconstruction is shown in Fig. \ref{fig:afm_recon}, where we see excellent correspondence between the original Morse force curve and the reconstructed polynomial.  

To make the inversion more realistic, we added Gaussian noise
corresponding to a maximum-spectral-amplitude-to-noise ratio $ \sim 10^{-4}$  (Fig. 3, inset), a dynamic range typical of the optical lever detectors in most AFMs.  With this noise we see that the polynomial miss-represents the force at maximum distance from the surface, where small changes in the coefficients $g_j$ with large $j$ have a large effect.   We also note that the polynomial may in general miss-represent the force near the point of maximum curvature, as many coefficients are needed to reconstruct sharp kinks in the force-distance curve.  To improve upon these results, we can carry the reconstruction one step further.  Using prior knowledge of the interaction in a particular experiment, for example that $F_{\rm ts}$ goes asymptotically to zero for large $\zeta$ and that a sharp kink exists at the contact point, one can make a weighted fit of the reconstructed polynomial to a particular force model containing far fewer than $j_{\rm{max}}$ parameters.   

Thus it is possible to extract a very nonlinear force-distance curve in the fast scanning mode \emph{without} using high frequency components of the cantilever response spectrum.  Intermodulation spectroscopy allows one to collect information in the form of coefficients $g_j$ with large $j$, by measuring high order IMPs at much lower frequencies than the $j^{\rm{th}}$ harmonic, where such information would occur with a single drive tone.  The ability of two drives to down-convert this information to lower frequencies, where it can be put near resonance and acquired with good transfer gain, is the major strength of the intermodulation spectroscopy technique.   
The method will be accurate if the dominant part of the spectral information is contained in the peaks which are used to construct the matrix $\mathbf{H}$, and if the traversed force curve is well approximated by $j_{\rm max}$ coefficients. Note that both amplitude and phase data are necessary to calculate the matrix $\mathbf{H}$.

We briefly note that we also successfully performed a simulated reconstruction of force-distance curves with two other drive schemes: One with a low-frequency drive and a drive at resonance, as sketched in Fig. 1, and another scheme with 
$\Omega_1=1/3$ and $\Omega_2=2/3+\Delta \Omega$. These drive schemes have the advantage that they produce IMPs of both odd and even order near resonance, and are therefore better with the highest $Q$ oscillators, and for nonlinearities containing both odd and even coefficients. 

In summary, intermodulation spectroscopy is a measurement technique which employs a high $Q$ oscillator and two drive frequencies to extract response caused by a nonlinear disturbance.  The drive frequencies are chosen commensurate with a greatest common divisor $\Delta \Omega$, and placed so as to produce many IMPs near resonance, where large transfer gain allows for detection with good signal-to-noise ratio.  A reconstruction algorithm was derived and
tested against simulated data including noise.   Excellent reconstruction of the nonlinearity was demonstrated with relatively few spectral peaks obtained at low frequencies.  
Together with the reconstruction algorithm, intermodulation spectroscopy should find wide-ranging use with the many types of sensors that are based on disturbance of a resonance. 

We acknowlegde helpful discussions with Jack Harris, Arvind Raman, and Ricardo Garcia. This work was supported by the Swedish Research Council and The Foundation for Strategic Research.

\bibliographystyle{mybst_PRL}
\bibliography{ref_imSpec}

\pagebreak

\end{document}